\documentclass[epj]{svjour}
\usepackage{graphics}

\begin{document}

\title{Scaling and self-averaging in the three-dimensional random-field Ising model}

\author{N.G. Fytas\thanks{e-mail: nfytas@phys.uoa.gr}
\and A. Malakis} \institute{Department of Physics, Section of
Solid State Physics, University of Athens, Panepistimiopolis, GR
15784 Zografos, Athens, Greece}

\date{Received: date / Revised version: date}

\abstract{We investigate, by means of extensive Monte Carlo
simulations, the magnetic critical behavior of the
three-dimensional bimodal random-field Ising model at the strong
disorder regime. We present results in favor of the two-exponent
scaling scenario, $\bar{\eta}=2\eta$, where $\eta$ and
$\bar{\eta}$ are the critical exponents describing the power-law
decay of the connected and disconnected correlation functions and
we illustrate, using various finite-size measures and properly
defined noise to signal ratios, the strong violation of
self-averaging of the model in the ordered phase.
}
\authorrunning{N.G. Fytas and A. Malakis} \titlerunning{Scaling and self-averaging in the three-dimensional random-field Ising model}

\maketitle

\section{Introduction}
\label{sec:1}

Phase transitions in pure systems~\cite{cardy-96} are already well
understood. The critical behavior of all physical quantities can
be described via critical exponents which are related through
(hyper)scaling relations to each other, so that only two
independent exponents remain. In contrast, phase transitions in
systems with (quenched) disorder~\cite{young-98} exhibit many
puzzles and are still far from being understood. In statistical
physics, and after the pioneering work of Imry and
Ma~\cite{imry-75}, the random-field Ising model (RFIM) is a widely
studied prototypical disordered system. It is
believed~\cite{fishman-79,cardy-84} to be in the same universality
class as the diluted antiferromagnet in a field, which can be
studied experimentally~\cite{belanger-98}.

Originally, it was believed~\cite{aharony-76,young-77,parisi-79}
that the critical behavior of the $d$-dimensional RFIM is equal to
that of the $d-2$ pure ferromagnet, implying that the $d=3$ RFIM
exhibits no ordered phase. This is of course not true, as has been
shown rigorously~\cite{bricmont-87} a few years later~\cite{note}.
In the meantime, a scaling
theory~\cite{villain-85,bray-85,fisher-86} for the RFIM was
developed, where the dimension $d$ has been replaced by $d-\theta$
in the hyperscaling relations, $\theta$ being a third independent
critical exponent, in contrast to the pure case. An alternative
approach~\cite{schwartz-85,schwartz-85b} leads to the consequence
that $\theta$ is not independent but related to the exponent
$\eta$ via $\theta=2-\eta$ (two-exponent scaling scenario).
Further evidence for the existence of only two independent
exponents was found by high-temperature
expansions~\cite{gofman-93}. This was mainly confirmed for the
RFIM with a Gaussian random-field distribution in $d=3$ by Monte
Carlo (MC) simulations~\cite{rieger-95b} and exact ground-state
calculations~\cite{hartmann-99,hartmann-01,middleton-02}, and in
$d=4$ also by ground-state calculations~\cite{hartmann-02}. The
first goal of the present paper is the numerical verification of
the above scenario for the case of the RFIM with a bimodal
random-field distribution, yet having in mind that the precision
of numerical simulations may never be good enough to actually
prove that this scenario is truly verified; rather we should see
this approach as a further test in favor of the proposed scaling
picture of the RFIM. A first attempt towards this direction has
been presented by Rieger and Young~\cite{rieger-93} in an early
paper, yet their analysis was restricted to rather small lattice
sizes ($L\leq 16$, $L$ being the linear lattice dimension).

The second part of our study focuses on another important relevant
issue in the study of disordered systems, that of self-averaging.
Although it has been known for many years now that for (spin and
regular) glasses there is no self-averaging in the ordered
phase~\cite{binder-86}, for random ferromagnets such a behavior
was first observed for the RFIM in a paper by Dayan \emph{et
al.}~\cite{dayan-93} and some years later for the random versions
of the Ising and Ashkin-Teller models by Wiseman and
Domany~\cite{WD-95}. These latter authors suggested a finite-size
scaling (FSS) ansatz describing the absence of self-averaging and
the universal fluctuations of random systems near critical points
that was refined and put on a more rigorous basis by the
intriguing renormalization-group work of Aharony and
Harris~\cite{AH-96}. Ever since, the subject of breakdown of
self-averaging is an important aspect in several theoretical and
numerical investigations of disordered spin
systems~\cite{EB-96,PSZ-97,WD-98,BF-98,TO-01,PS-02,BC-04,MG-05,fytas-06,wu-06,GL-07}.
In fact, most recently, Efrat and Schwartz~\cite{efrat-06} showed
that the property of lack of self-averaging in disordered systems
may be turned into a useful tool that can provide an independent
measure to distinguish between the ordered and disordered phases
of the system. In view of this increasing interest in the
understanding of the self-averaging properties of disordered
systems we attempt in the present paper to examine and apply the
theoretical predictions to the RFIM, a suitable candidate which is
already known to suffer from a strong violation of self-averaging
in terms of the thermal properties~\cite{PS-02,fytas-06}.

The Hamiltonian describing the RFIM is
\begin{equation}
\label{eq:1}
\mathcal{H}=-J\sum_{<i,j>}S_{i}S_{j}-h\sum_{i}h_{i}S_{i},
\end{equation}
where $S_{i}$ are Ising spins, $J>0$ is the nearest-neighbors
ferromagnetic interaction, $h$ is the disorder strength, also
called randomness of the system, and $h_{i}$ are independent
quenched random fields obtained here from a bimodal distribution
of the form
\begin{equation}
\label{eq:2} P(h_{i})=\frac{1}{2}[\delta(h_{i}-1)+\delta(h_{i}+1].
\end{equation}
The critical value of $h$, denoted as critical disorder strength
$h_{c}$, above which no phase transition occurs in the
thermodynamic limit is known with good accuracy for the bimodal
RFIM to be $h_{c}=2.21(1)$~\cite{hartmann-99,fytas-08}. In
particular, in the present paper we consider the model defined in
equations~(\ref{eq:1}) and (\ref{eq:2}) in the strong disorder
regime ($h\in [1.9, 2.25]$) and embedded in simple cubic lattices
with periodic boundary conditions and $N=L^{3}$ spins, where the
linear dimension $L$ takes values in the range $L=4-32$. The
ensembles of random-field realizations $Q$ $(q=1,2,\ldots,Q)$
simulated for the present study are the following: $Q_{L\leq
16}=1000$ and $Q_{L> 16}=500$. The total computer time used for
the simulations was several ($\sim 8$) Intel Pentium-Pro months.

The rest of the paper is laid out as follows: In
Section~\ref{sec:2} we briefly describe here our numerical
implementation, that is based on an efficient implementation of
the Wang-Landau (WL) method~\cite{wang-01}, as also presented in a
number of papers dealing with the simulation aspects of disordered
systems~\cite{malakis-04,fytas-08b,malakis-09}. In
Section~\ref{sec:3} we discuss the FSS behavior of the magnetic
quantities of the model and provide evidence in favor of the
two-exponent scaling scenario. Subsequently, in
Section~\ref{sec:4} we investigate the self-averaging properties
of the model in terms of broad distributions and we illustrate
using various finite-size measures and properly defined noise to
signal ratios, the strong violation of self-averaging of the model
in the ordered phase. Finally, we summarize our conclusions in
Section~\ref{sec:5}.

\section{Numerical approach}
\label{sec:2}

MC simulations in statistical physics now have a history of nearly
half a century starting with the seminal work of Metropolis. While
the Metropolis algorithm has been established as the standard
algorithm for importance sampling it suffers from two problems:
the inability to directly calculate the partition function, free
energy or entropy, and critical slowing down near phase
transitions and in disordered systems. In a standard MC algorithm
a series of configurations is generated according to a given
distribution, usually the Boltzmann distribution in classical
simulations. While this allows the calculation of thermal
averages, it does not give the partition function, nor the free
energy. They can only be obtained with limited accuracy as a
temperature integral of the specific heat, or by using maximum
entropy methods. The problem of critical slowing down has been
overcome for second-order phase transitions by cluster update
schemes~\cite{wang}. For first-order phase transitions and systems
with rough free-energy landscapes (such the present model under
study) a decisive improvement was achieved some years ago via the
WL algorithm~\cite{wang-01}.

In the last few years our group has used an entropic sampling
implementation of the WL algorithm~\cite{wang-01} to
systematically study some simple~\cite{malakis-04}, but also some
more complex
systems~\cite{fytas-06,fytas-08,fytas-08b,malakis-09}. One basic
ingredient of this implementation is a suitable restriction of the
energy subspace for the implementation of the WL algorithm. This
was originally termed as the critical minimum energy subspace
(CrMES) restriction~\cite{malakis-04} and it can be carried out in
many alternative ways, the simplest being that of observing the
finite-size behavior of the tails of the energy probability
density function (e-pdf) of the system~\cite{malakis-04}.

Complications that may arise in complex systems, i.e. random
systems or systems showing a first-order phase transition, can be
easily accounted for by various simple modifications that take
into account possible oscillations in the e-pdf and expected
sample-to-sample fluctuations of individual realizations. In our
recent papers~\cite{fytas-08,fytas-08b,malakis-09}, we have
presented details of various sophisticated routes for the
identification of the appropriate energy subspace $(E_{1},E_{2})$
for the entropic sampling of each realization. In estimating the
appropriate subspace from a chosen pseudocritical temperature one
should be careful to account for the shift behavior of other
important pseudocritical temperatures and extend the subspace
appropriately from both low- and high-energy sides in order to
achieve an accurate estimation of all finite-size anomalies. Of
course, taking the union of the corresponding subspaces, insures
accuracy for the temperature region of all studied pseudocritical
temperatures.

The up to date version of our implementation uses a combination of
several stages of the WL process. First, we carry out a starting
(or preliminary) multi-range (multi-R) stage, in a very wide
energy subspace. This preliminary stage is performed up to a
certain level of the WL random walk. The WL refinement is
$G(E)\rightarrow f\cdot G(E)$, where $G(E)$ is the density of
states (DOS) and we follow the usual modification factor
adjustment $f_{j+1}=\sqrt{f_{j}}$ and $f_{1}=e$. The preliminary
stage may consist of the levels : $j=1,\ldots,j=18$ and to improve
accuracy the process may be repeated several times. However, in
repeating the preliminary process and in order to be efficient, we
use only the levels $j=13,\ldots,18$ after the first attempt,
using as starting DOS the one obtained in the first random walk at
the level $j=12$. From our experience, this practice is almost
equivalent to simulating the same number of independent WL random
walks. Also in our recent studies we have found out that is much
more efficient and accurate to loosen up the originally applied
very strict flatness criteria~\cite{malakis-04}. Thus, a variable
flatness process starting at the first levels with a very loose
flatness criteria and assuming at the level $j=18$ the original
strict flatness criteria is now days used. After the above
described preliminary multi-R stage, in the wide energy subspace,
one can proceed in a safe identification of the appropriate energy
subspace using one or more alternatives outlined in
reference~\cite{malakis-04}.

The process continues in two further stages (two-stage process),
using now mainly high iteration levels, where the modification
factor is very close to unity and there is not any significant
violation of the detailed balance condition during the WL process.
These two stages are suitable for the accumulation of histogram
data (for instance energy-magnetization histograms), which can be
used for an accurate entropic calculation of non-thermal
thermodynamic parameters, such as the order parameter and its
susceptibility~\cite{malakis-04}. In the first (high-level) stage,
we follow again a repeated several times (typically $\sim 5-10$)
multi-R WL approach, carried out now only in the restricted energy
subspace. The WL levels may be now chosen as $j=18,19,20$ and as
an appropriate starting DOS for the corresponding starting level
the average DOS of the preliminary stage at the starting level may
be used. Finally, the second (high-level) stage is applied in the
refinement WL levels $j=j_{i},\ldots,j_{i}+3$ (typically
$j_{i}=21$), where we usually test both an one-range (one-R) or a
multi-R approach with large energy intervals. In the case of the
one-R approach we have found very convenient and in most cases
more accurate to follow the Belardinelli and
Pereyra~\cite{belardinelli-07} adjustment of the WL modification
factor according to the rule $\ln f\sim t^{-1}$, where $t$ refers
to the MC time of the process. Finally, it should be also noted
that by applying in our scheme a separate accumulation of
histogram data in the starting multi-R stage (in the wide energy
subspace) offers the opportunity to inspect the behavior of all
basic thermodynamic functions in an also wide temperature range
and not only in the neighborhood of the finite-size anomalies.

Closing this review on the numerical part of this work, we would
like to point out that, the above scheme may also be extended to
accumulate multi-parametric histograms, and provide information
for the critical properties of the system that are not easily
accessible with other methodologies. For more details on this
extension we refer the reader to reference~\cite{fytas-08}.

\section{Two-exponent scaling}
\label{sec:3}

We start by studying the critical behavior of the susceptibility
$\chi$ and disconnected susceptibility $\chi_{dis}$, which are
given, in general, by the standard expressions
\begin{equation}
\label{eq:3} \chi=\frac{1}{NT}\sum_{i,j=1}^{N}(\langle
S_{i}S_{j}\rangle-\langle S_{i}\rangle \langle S_{j} \rangle),
\end{equation}
and
\begin{equation}
\label{eq:4} \chi_{dis}=\frac{1}{N}\sum_{i,j=1}^{N}\langle
S_{i}\rangle \langle S_{j} \rangle,
\end{equation}
where $\langle \ldots \rangle$ denotes a thermal average. We now
follow the scaling arguments based on the droplet
theory~\cite{villain-85,bray-85,fisher-86}, as presented by Dayan
\emph{et al.}~\cite{dayan-93}. According to these arguments, most
samples in a large ensemble of realizations of the random field
have only one thermally populated minimum. For these samples
(called hereafter as typical) the two terms in
equation~(\ref{eq:3}) almost cancel  at $T_{c}$, and while each
term has separately the divergence of $\chi_{dis}$, i.e.
\begin{equation}
\label{eq:5} \chi_{dis}\sim L^{4-\bar{\eta}}\;\;\;
(4-\bar{\eta}=\bar{\gamma}/\nu),
\end{equation}
their difference is given by
\begin{equation}
\label{eq:6} \chi_{t}\sim L^{2-\eta}\;\;\; (2-\eta=\gamma/\nu),
\end{equation}
with probability $p_{t}\sim 1$ (the subscript $t$ stands for
typical). The cancellation of the most divergent terms in the
expression of the susceptibility is rigorously true on average,
according to the Schwartz-Soffer inequality~\cite{schwartz-85}.
Thus, following reference~\cite{schwartz-85}, one shows the
inequality $4-\bar{\eta}\leq d-2\beta/\nu$, and assuming that this
holds as an equality (see also reference~\cite{EB-96}) one derives
the modified hyperscaling relation
\begin{equation}
\label{eq:7} 4-\bar{\eta}=d-2\beta/\nu,
\end{equation}
which can be also written in the form
\begin{equation}
\label{eq:8} (d-\theta)\nu=2\beta+\gamma.
\end{equation}
The exponent $\theta$ in the above equation accounts for the
violation of the hyperscaling relation and is given, according to
the above equations~(\ref{eq:7}) and (\ref{eq:8}), by
\begin{equation}
\label{eq:9} \theta=2-\bar{\eta}+\eta.
\end{equation}

\begin{figure}
\resizebox{1 \columnwidth}{!}{\includegraphics{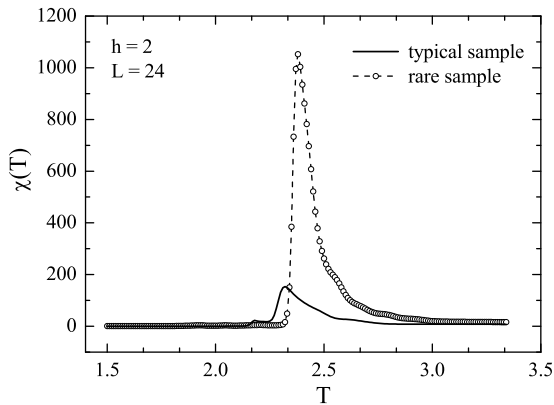}}
\caption{Magnetic susceptibility as a function of temperature of a
typical (solid line) and a rare (dotted line) sample for $h=2$ and
$L=24$.} \label{fig:1}
\end{figure}
Furthermore, rare samples with more than one minimum (thermally
populated) may appear with a probability $p_{r}\sim L^{-\theta}$
(where the subscript $r$ stands for rare) and, for these samples,
the two terms in equation~(\ref{eq:3}) do not cancel at $T_{c}$,
and their difference is of the same order of each term, i.e.
\begin{equation}
\label{eq:10} \chi_{r}\sim L^{4-\bar{\eta}}.
\end{equation}
Although with increasing $L$ the probability of obtaining one of
these rare samples decreases, the susceptibility of a rare sample
gets much larger, relatively to the value of a typical sample
($\chi_{r}\gg \chi_{t}$, see Figure~\ref{fig:1}). Thus, we need to
average over large ensembles of random fields in order to obtain
good statistics.

Averaging over $Q$ realizations of the random field and using
equations~(\ref{eq:6}) and (\ref{eq:10}), we see that for the
first moment $[\chi]_{av}$ (where $[\ldots]_{av}$ denotes
randomness averaging), both typical and rare samples give a
comparable contribution, so that at $T_{c}$
\begin{equation}
\label{eq:11} [\chi]_{av}=\sum_{i=t,r}p_{i}\chi_{i}\sim 1\cdot
L^{2-\eta}+L^{-\theta}\cdot L^{4-\bar{\eta}}\sim L^{2-\eta}.
\end{equation}
The behavior of the average disconnected susceptibility at $T_{c}$
is, according to equation~(\ref{eq:5}),
\begin{equation}
\label{eq:12} [\chi_{dis}]_{av}\sim L^{4-\bar{\eta}}.
\end{equation}

At this point we shall assume that the above power laws are valid
also for the pseudocritical temperatures of some relevant
thermodynamic quantities. In particular, we will consider here
(mainly) the sequence of pseudocritical temperatures that
correspond to the maxima (for different $L$) of the average
susceptibility ($[\chi]_{av}^{\ast}$) defined as
\begin{equation}
\label{eq:13}
[\chi]_{av}^{\ast}=\max_{T}\{[\chi]_{av}\}=\max_{T}\{\frac{1}{Q}\sum_{q=1}^{Q}\chi_{q}(T)\},
\end{equation}
where the subscript $q$ refers to a particular random realization
of the quenched disorder and $\chi_{q}$ to the corresponding
magnetic susceptibility. These pseudocritical temperatures will be
denoted in the sequel as $T_{L,[\chi]_{av}^{\ast}}$ (for the case
of the specific heat the corresponding pseudocritical temperatures
will be denoted as $T=T_{L,[C]_{av}^{\ast}}$).
\begin{figure}
\resizebox{1 \columnwidth}{!}{\includegraphics{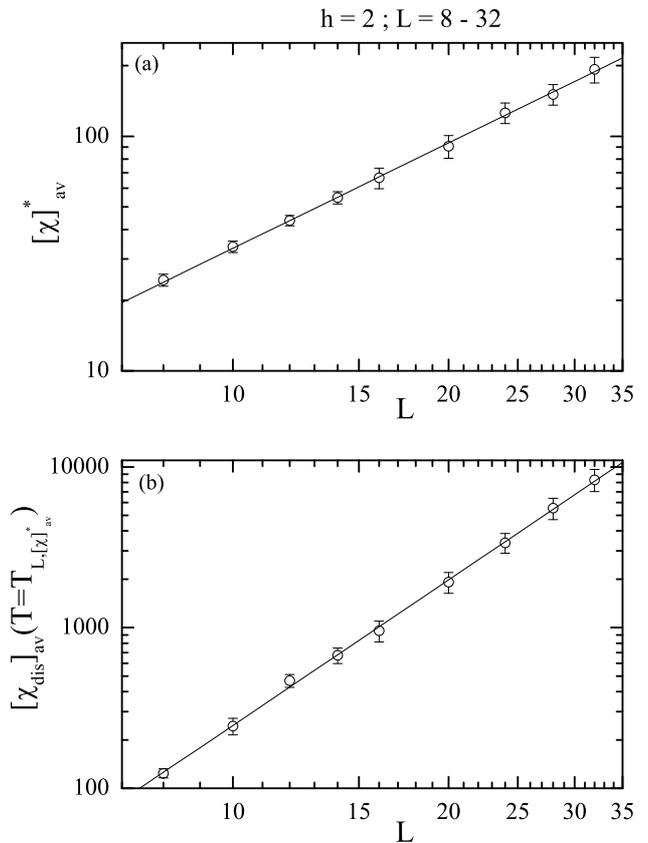}}
\caption{FSS in a double-logarithmic scale of (a) the maxima of
the average susceptibility and (b) the average disconnected
susceptibility at the temperature where the average susceptibility
attains its maximum, for $h=2$ and $L=8-32$. In both panels the
solid lines show linear fitting attempts.} \label{fig:2}
\end{figure}
In Figure~\ref{fig:2}(a) we plot the maxima of the average
susceptibility versus the lattice size $L$ in a log-log plot for
the case $h=2$. The full line is a least-squares straight-line
fitting which gives a slope of $1.49(2)$, that is according to
equation~(\ref{eq:11}) an estimate for $2-\eta$, yielding
\begin{equation}
\label{eq:14} \eta=0.51(2).
\end{equation}
The average disconnected susceptibility at
$T_{L,[\chi]_{av}^{\ast}}$, denoted as $[\chi_{dis}]_{av}^{\ast}$,
is depicted in Figure~\ref{fig:2}(b), also for $h=2$, and is
expected to obey the power law~(\ref{eq:12}). Using again a
least-squares fitting of the data to a straight line in the
log-log plot we get the estimate for $4-\bar{\eta}$ to be
$2.99(3)$, and therefore
\begin{equation}
\label{eq:15} \bar{\eta}=1.01(3).
\end{equation}
From equations~(\ref{eq:14}) and (\ref{eq:15}) we see that the
Schwartz-Soffer inequality $\bar{\eta}\leq
2\eta$~\cite{schwartz-85} is fulfilled within error bars as an
equality, supporting the two-exponent scaling scenario, i.e.
$\bar{\eta}=2\eta$~\cite{gofman-93,schwartz-85,schwartz-85b}. The
exponent $\theta$ is estimated from equation~(\ref{eq:9}) to be
\begin{equation}
\label{eq:16} \theta=1.50(3).
\end{equation}

Overall, the estimated values for $\eta$, $\bar{\eta}$, and
$\theta$ are in very close agreement with the values
\begin{equation}
\label{eq:17} \eta=0.5\;\;\;\;\;\bar{\eta}=1\;\;\;\;\;\theta=1.5,
\end{equation}
that can be predicted from elementary considerations in the
ordered phase (see also the discussion in
reference~\cite{dayan-93}) and agree with the early dimensional
reduction prediction $d\rightarrow d/2$ of
Shapir~\cite{shapir-85}. Moreover, the resulting value for the
exponent $\beta$ is very close to zero, as already noted by Dayan
\emph{et al.}~\cite{dayan-93}. The resulting very small value for
$\beta$ ($\beta=0.007(5)$) with a relatively large error could be
compared with renormalization group results
($\beta=0.0200(5)$)~\cite{falicov-95} and zero temperature studies
($\beta=0.017(5)$)~\cite{middleton-02}.

Up to this point, our analysis and discussion followed the
assumption of a continuous phase transition, which is actually the
most widely accepted scenario in the random-field community.
However, the estimated value for the critical exponent $\beta$
raises some doubts of whether this is actually true, especially
for the present case of the bimodal RFIM, for which mean-field
theory predicted a tricritical point at high values of the random
field~\cite{aharony-78}. This main issue has regained interest
after the recent observations, in both the
bimodal~\cite{hernandez-97,hernandez-08} and
Gaussian~\cite{wu-06,martin} cases, of first-order-like features
at the strong disorder regime. In particular first-order-like
features, such as the appearance of the characteristic double-peak
structure of the canonical e-pdf, have been recently reported for
both the Gaussian and the bimodal distributions of the $d=3$ RFIM.

However, in a recent paper Fytas et al.~\cite{fytas-08b} presented
a detailed study of the first-order transition features - in terms
of the surface tension and latent heat - of the $d=3$ RFIM, having
in mind that a mere observation of a first-order structure is not
sufficient for the identification of the transition. As it was
shown in reference~\cite{fytas-08b}, this is especially true for
the RFIM, since its critical behavior is obscured by strong and
complex finite-size effects, involving also the important issue of
the lack of self-averaging, discussed here in the next Section.
The results of reference~\cite{fytas-08b} for two values of the
disorder strength in the strong disorder regime ($h=2$ and
$h=2.25$) clearly indicated that the interface tension vanishes
and the two peaks of the e-pdf move together in the thermodynamic
limit and therefore provided convincing evidence that the
transition is continuous and that there in no tricritical point
along the phase transition line. These results pointed to an
unconventional continuous transition, in which the e-pdf
approaches two delta functions that move together in the
thermodynamic limit. Such an unconventional behavior has been
originally predicted by Eichhorn and Binder~\cite{EB-96}, for the
the order-parameter pdf of the $d=3$ random-field three-state
Potts model and found its theoretical justification in the
framework of a refined FSS theory a few months ago in a profound
paper by Vink et al.~\cite{vink}.

Closing this Section, we may also calculate the critical exponents
$\gamma$ of the susceptibility and $\bar{\gamma}$ of the
disconnected susceptibility
\begin{eqnarray} \label{eq:18}
\gamma&=&\nu(2-\eta)=1.95(27)\nonumber
\\ \bar{\gamma}&=&\nu(4-\bar{\eta})=3.92(54),
\end{eqnarray}
which also compare favorably with previous
estimations~\cite{dayan-93,rieger-93,gofman-93,rieger-95b,hartmann-01}.
Note that in equation~(\ref{eq:18}) the large error bounds come
mainly from the error bounds in $\nu$, since we have used the
value $\nu=1.31(18)$ estimated in our previous
work~\cite{fytas-06}. We would like to note here that, an
analogous study of $\chi$ and $\chi_{dis}$ was performed by
Eichhorn and Binder~\cite{EB-96} for the $d=3$ three-state
random-field Potts model. Their findings provided also qualitative
evidence for the two-exponent scaling description of random-field
systems.

\section{Broad distributions and lack of self-averaging}
\label{sec:4}

\begin{figure}
\resizebox{1 \columnwidth}{!}{\includegraphics{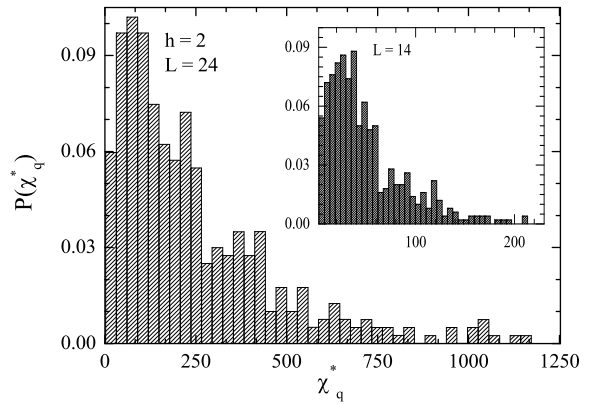}}
\caption{The histograms for the probability distribution
$P(\chi_{q}^{\ast})$ of the susceptibility maxima for $L=24$ (and
$L=14$ in the inset).} \label{fig:3}
\end{figure}
Let us start this Section by recapitulating some basic definitions
and concepts on self-averaging. As discussed above, our numerical
studies of disordered systems are carried out near their critical
points using finite samples; each sample $q$ is a particular
random realization of the quenched disorder. A measurement of a
thermodynamic property $w$ yields a different value for the exact
thermal average $w_{q}$ of every sample $q$. In an ensemble of
disordered samples of linear size $L$ the values of $w_{q}$ are
distributed according to a probability distribution $P(w_{q})$.
The behavior of this distribution is directly related to the issue
of self-averaging. In particular, by studying the behavior of the
width of $P(w_{q})$ with increasing the system size $L$, one may
address qualitatively the issue of self-averaging, as has already
been stressed by previous authors~\cite{WD-98}. In general, we
characterize the distribution $P(w_{q})$ by its average $[w]_{av}$
and also by the relative variance
\begin{equation}
\label{eq:19}
R_{w}=\frac{V_{w}}{[w]_{av}^{2}}=\frac{[w^{2}]_{av}-[w]_{av}^{2}}{[w]_{av}^{2}}.
\end{equation}
Suppose now that $w$ is a singular density of an extensive
thermodynamic property, such as $M$ or $\chi$, or the singular
part of $e(=E/N)$ and $C$. The system is said to exhibit
self-averaging if $R_{w}\rightarrow 0$ as $L\rightarrow \infty$.
If $R_{w}$ tends to a non-zero value, i.e $R_{w}\rightarrow const
\neq 0$ as $L\rightarrow \infty$, then the system exhibits lack of
self-averaging.

The importance of the above concepts has been illustrated by
Aharony and Harris~\cite{AH-96} and their main conclusions are
summarized as follows: (i) Outside the critical temperature:
$R_{w}=0$. In a finite geometry, the correlation length $\xi$ is
finite for $T\neq T_{c}$ and it can be found, using general
statistical arguments, originally introduced by
Brout~\cite{brout-59}, that $R_{w}\propto (\xi/L)^{d}\rightarrow
0$, as $L\rightarrow \infty$. This is called strong
self-averaging. (ii) At the critical temperature there exist two
possible scenarios: (a) models in which according to the Harris
criterion~\cite{harris-74} the disorder is relevant
($\alpha_{p}>0$): $R_{w}\neq 0$. Then, the system at the critical
point is not self-averaging and (b) models in which according to
the Harris criterion disorder is irrelevant ($\alpha_{p}<0$):
$R_{w}=0$. In this case $R_{w}$ scales as $L^{\alpha/\nu}$, where
$\alpha$ and $\nu$ are the critical exponents of the pure system,
which are the same in the disordered one.
\begin{figure}
\resizebox{1 \columnwidth}{!}{\includegraphics{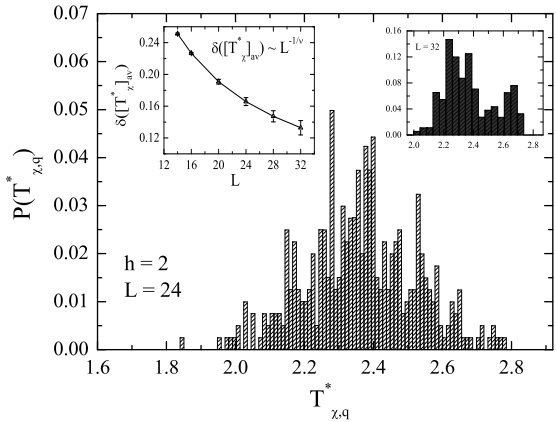}}
\caption{The $h=2$ histograms of the probability distribution
$P(T_{\chi,q}^{\ast})$ for $L=24$ (main panel) and $L=32$ (right
inset). The left inset shows the FSS of the sample-to-sample
variance of the average $[T_{\chi}^{\ast}]_{av}$.} \label{fig:4}
\end{figure}
This is called weak self-averaging. (iii) The pseudocritical
temperatures $T_{w,q}^{\ast}$ of the disordered system are
distributed with a width $\delta ([T_{w}^{\ast}]_{av})$ which
scales with the system size as $\delta ([T_{w}^{\ast}]_{av})\sim
L^{-n}$, where $n=d/2$ or $n=1/\nu$, depending on whether the
disordered system is controlled by the pure or the random fixed
point, respectively. The above behavior is now well established by
the pioneering works of Aharony and Harris~\cite{AH-96} and
Wiseman and Domany~\cite{WD-95,WD-98}.

We start the presentation of our results with the probability
distribution of the susceptibility maxima $P(\chi^{\ast}_{q})$.
$P(\chi^{\ast}_{q})$ is expected to be a broad distribution with a
most probable value of the order $L^{2-\eta}$ and a tail extending
to much larger values of the order
$L^{4-\bar{\eta}}$~\cite{dayan-93}. Figure~\ref{fig:3} and the
corresponding inset show the normalized histograms of
$P(\chi^{\ast}_{q})$ at $h=2$ for lattice sizes $L=24$ and $L=14$,
respectively.
\begin{figure}
\resizebox{1 \columnwidth}{!}{\includegraphics{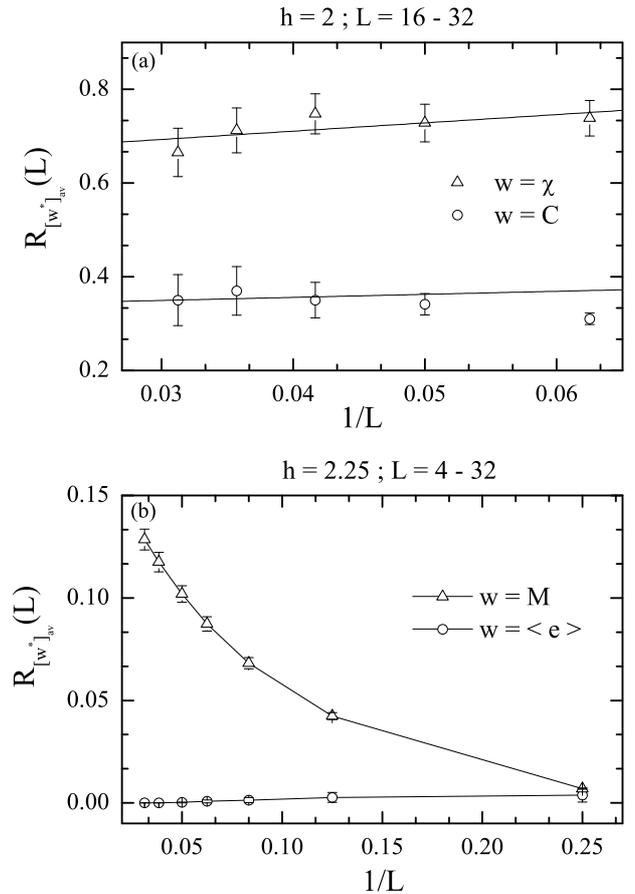}}
\caption{FSS behavior of the ratio $R_{[w^{\ast}]_{av}}(L)$
defined in equation~(\ref{eq:22}) for (a) the specific heat $w=C$
and magnetic susceptibility $w=\chi$ for $h=2$ and (b) the
magnetization $w=M$ and the mean energy per spin $w= \langle e
\rangle$ for $h=2.25$. In panel (a) the solid lines show linear
fitting extrapolations to the limit $L \rightarrow \infty$.}
\label{fig:5}
\end{figure}
In both cases $P(\chi^{\ast}_{q})$ has its peak at a rather small
value of $\chi^{\ast}_{q}$ but there is a clearly developing long
tail extending to much larger values of $\chi^{\ast}_{q}$. As
expected, the tail is longer for the larger lattice size.
Subsequently, in the main panel and right inset of
Figure~\ref{fig:4} we plot the $h=2$ histograms for the
probability distribution $P(T_{\chi,q}^{\ast})$ of the
pseudocritical temperatures of the magnetic susceptibility for
$L=24$ and $L=32$, respectively. The left inset in
Figure~\ref{fig:4} illustrates the scaling of the sample-to-sample
variance of the sample average
\begin{equation}
\label{eq:20}
[T_{\chi}^{\ast}]_{av}=(1/Q)\sum_{q=1}^{Q}T_{\chi,q}^{\ast}.
\end{equation}
Assuming that the width of these sample-to-sample fluctuations
scales with the linear size $L$ according to
\begin{equation}
\label{eq:21} \delta([T_{\chi}^{\ast}]_{av})\sim L^{-1/\nu},
\end{equation}
we obtain, from the very good fitting for the sizes $L=14-32$
shown in the inset of Figure~\ref{fig:4}, the value $\nu=1.30(2)$.
This estimate is in excellent agreement with the value
$\nu=1.31(18)$ used in equation~(\ref{eq:18}) and estimated in our
previous paper~\cite{fytas-06} from the scaling of the specific
heat's pseudocritical temperature and consists further evidence
for the breakdown of self-averaging in the RFIM.

We turn now to study the behavior of the ratio $R_{w}$ defined
above. In our formalism, we may define the variance ratio $R_{w}$
in two distinct forms. One $L$-dependent ratio for the sample
average of the individual maxima of $w$
$[w^{\ast}]_{av}=(1/Q)\sum_{q=1}^{Q}w_{q}^{\ast}$
\begin{equation}
\label{eq:22} R_{[w^{\ast}]_{av}}(L)=
\frac{V_{[w^{\ast}]_{av}}}{[w^{\ast}]_{av}^{2}},
\end{equation}
and one $T$-dependent for the average curve $[w]_{av}$
\begin{equation}
\label{eq:23} R_{[w]_{av}}(T)=\frac{V_{[w]_{av}}}{[w]_{av}^{2}}.
\end{equation}
In Figure~\ref{fig:5}(a) the behavior of the ratio
$R_{[w^{\ast}]_{av}}(L)$, defined in equation~(\ref{eq:22}), where
$w^{\ast}=\chi^{\ast}=\chi(T=T_{L,[\chi]_{av}^{\ast}})$ and
$w^{\ast}=C^{\ast}=C(T=T_{L,[C]_{av}^{\ast}})$ is illustrated for
$L=16-32$, as a function of the inverse linear size for disorder
strength $h=2$. The straight lines are extrapolations to
$L\rightarrow \infty$. In both cases a clear saturation to a
non-zero limiting value is observed, i.e. $\lim_{L\rightarrow
\infty}R_{[\chi^{\ast}]_{av}}(L)=0.64(9)$ and $\lim_{L\rightarrow
\infty}R_{[C^{\ast}]_{av}}(L)=0.33(5)$, indicating violation of
self-averaging~\cite{fytas-06,AH-96,WD-98,efrat-06}.
\begin{figure}
\resizebox{1 \columnwidth}{!}{\includegraphics{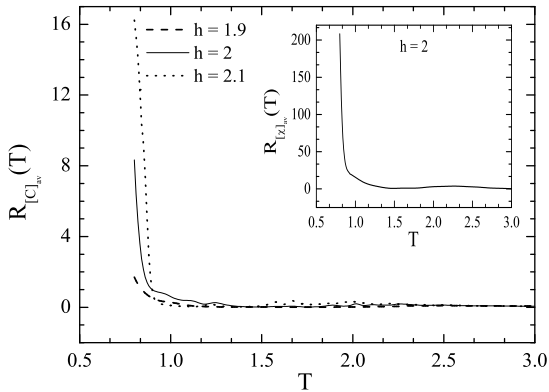}}
\caption{Low-temperature behavior of the ratio $R_{[w]_{av}}(T)$
defined in equation~(\ref{eq:23}) of the specific heat ($w=C$) and
$h=1.9,2,$ and $2.1$ in the main panel and the magnetic
susceptibility ($w=\chi$) and $h=2$ in the corresponding inset.
The linear size dimension is $L=10$.} \label{fig:6}
\end{figure}
For the case of the susceptibility this violation is much
stronger. Figure~\ref{fig:5}(b) presents the behavior of the ratio
$R_{[w^{\ast}]_{av}}(L)$, where now
$w^{\ast}=M^{\ast}=M(T=T_{L,[\chi]_{av}^{\ast}})$ and
$w^{\ast}=\langle e\rangle^{\ast}=\langle
e\rangle(T=T_{L,[C]_{av}^{\ast}})$ as a function of the inverse
linear size for sizes $L=4-32$ and disorder strength $h=2.25$. It
is clear that, while for the magnetization the ratio
$R_{[M^{\ast}]_{av}}(L)$ approaches a non zero value in the limit
$L\rightarrow \infty$ and therefore violation of self-averaging is
expected, the same ratio for the mean energy $R_{[\langle
e\rangle^{\ast}]_{av}}(L)$ starts with some very small values but
finally approaches zero with increasing lattice size, indicating
that in the infinite size limit the mean energy of the system is
self-averaging.

Finally, in Figure~\ref{fig:6} we present the low-temperature
behavior of the ratio $R_{[w]_{av}}(T)$, defined in
equation~(\ref{eq:23}), for the specific heat $w=C$ and the
susceptibility $w=\chi$ (inset) at $h=2$ and for a lattice size
$L=10$. The ratio $R_{[C]_{av}}$ for $h=1.9$ and $h=2.1$ is also
shown in the main panel together with that of $h=2$, calculated
via the extrapolation scheme described in
reference~\cite{fytas-08}. Our intention is to identify the
temperature variation of the non self-averaging property of the
average specific heat and susceptibility. We observe that as we
move from the disordered phase to the ordered phase this ratio
increases, and gets its maximum value for the lowest temperatures
shown. This indicates strong sample-to-sample fluctuations and non
self-averaging behavior (which is more pronounced with increasing
disorder strength), it becomes more evident as we enter the
ordered phase, and is in agreement with previous
studies~\cite{dayan-93,rieger-93,rieger-95b}. Since the
sample-averaged pseudocritical temperature for $L=10$ is of the
order of $[T_{\chi}^{\ast}]_{av}\simeq 2.7$, the violation of
self-averaging goes well into the ordered phase, and this should
be compared with Figure 3 of reference~\cite{efrat-06}.

Let us comment here that, as stated above, possible breakdown of
self-averaging may be traced back to situations where the
correlation length $\xi$ becomes of the order of the linear size
of the system. The usual state of affairs is that the correlation
length diverges just on the boundary between the ordered and
disordered phase and consequently any break down of self-averaging
may be observed only in the vicinity of the
boundary~\cite{fytas-06,AH-96,WD-98,efrat-06}. However, as Dayan
et al.~\cite{dayan-93} showed, the situation in the RFIM is quite
different: the correlation length is of order of the linear size
of the system everywhere in the ordered phase and not just at the
boundary of the phase, indicating a violation of self-averaging in
the ordered phase, as clearly manifested in our
Figure~\ref{fig:6}. Of course, self-averaging is expected to be
restored at high-temperatures (above $T_{c}$) that $\xi\ll L$,
because in this case, the system can be divided into independent
regions of size $\xi$ and any measurement on the whole sample will
be an average on these regions. This latter behavior is also
depicted in our Figure~\ref{fig:6}, where the ratio decreases
rapidly and approaches zero for high temperatures for both the
specific heat and the susceptibility.

\section{Conclusions}
\label{sec:5}

In the present paper we investigated numerically some important
aspects of the critical properties of the $d=3$ RFIM with a
bimodal random-field distribution in the strong disorder regime.
In the first part of our study we focused on the scaling aspects
of the model and presented evidence in favor of the so-called
two-exponent scaling scenario. We also derived accurate values for
the magnetic critical exponents $\beta$, $\gamma$, and
$\bar{\gamma}$ of the model in good agreement with the existing
estimates in the literature.

In the second part, we investigated the self-averaging properties
of the model in terms of probability distributions of certain
thermodynamic quantities and we illustrated, using various
finite-size measures and properly defined noise to signal ratios,
the strong violation of self-averaging of the model in the ordered
phase. Additionally, a FSS analysis of the width of the
distribution of the sample-dependent pseudocritical temperatures
of the magnetic susceptibility verified the theoretical
expectations of Aharony and Harris and provided an alternative
approach of extracting the value of the correlation length's
exponent $\nu$. Overall, the data and analysis presented in this
paper indicate that the most complicated issue of breakdown of
self-averaging in disordered systems deserves special attention
and can be easily transformed into a useful tool that constitutes
an alternative approach to criticality.

\end{document}